\documentclass{PoS}

\usepackage{epsfig}
\usepackage{xspace}
\usepackage{amsmath}  
\usepackage{relsize}  
\usepackage{verbatim}

\newcommand{\etal}{\textit{et al.\@}\xspace}
\newcommand{\commenthww}[1]{}

\def\q2{\ensuremath{q^2}\xspace}
\def\mES{\ensuremath{m_{\rm ES}}\xspace}
\def\DeltaE{\ensuremath{\Delta E}\xspace}
\def\Bzdslnu{\ensuremath{B^{0} \rightarrow D^{*-} \ell^+\nu}\xspace}
\def\Bzpilnu{\ensuremath{B^{0} \rightarrow \pi^-\ell^+\nu}\xspace}
\def\Bppizlnu{\ensuremath{B^{+} \rightarrow \pi^0\ell^+\nu}\xspace}
\def\Bzrholnu{\ensuremath{B^{0} \rightarrow \rho^-\ell^+\nu}\xspace}
\def\Bprhozlnu{\ensuremath{B^{+} \rightarrow \rho^0\ell^+\nu}\xspace}
\def\Bpilnu{\ensuremath{B \rightarrow \pi \ell\nu}\xspace}
\def\Brholnu{\ensuremath{B \rightarrow \rho \ell\nu}\xspace}
\def\BXulnu{\ensuremath{B \rightarrow X_u\ell\nu}\xspace}
\def\BXclnu{\ensuremath{B \rightarrow X_c\ell\nu}\xspace} 
\def\BRBzpilnu{\ensuremath{{\cal B}(B^{0} \rightarrow \pi^-\ell^+\nu})\xspace}
\def\BRBzrholnu{\ensuremath{{\cal B}(B^{0} \rightarrow \rho^-\ell^+\nu})\xspace}
\def\qq{\ensuremath{q\bar{q}\xspace}}
\def\Vub  {\ensuremath{|V_{ub}|}\xspace}
\def\babar{\mbox{\slshape B\kern-0.1em{\smaller A}\kern-0.1em
    B\kern-0.1em{\smaller A\kern-0.2em R}}\xspace}
\def\BBbar {\ensuremath{B\overline B}\xspace}
\def\epem  {\ensuremath{e^+e^-}\xspace}
\mathchardef\Upsilon="7107
\def\Y#1S{\ensuremath{\Upsilon{(#1S)}}\xspace}
\def\FourS {\Y4S}
\def\invfb   {\ensuremath{\mbox{\,fb}^{-1}}\xspace}
\newcommand{\mev}{\ensuremath{\mathrm{\,Me\kern -0.1em V}}\xspace}
\newcommand{\gev}{\ensuremath{\mathrm{\,Ge\kern -0.1em V}}\xspace}

\def\dBdq2{\ensuremath{\Delta{\cal B}/\Delta q^2}\xspace}

\title{Study of \Bpilnu and \Brholnu decays and 
  determination of \Vub at \babar}

\ShortTitle{$B \rightarrow (\pi/\rho)\ell\nu$ and \Vub at \babar}

\author{\speaker{H. Wells Wulsin} 
  \thanks{Representing the \babar Collaboration.}
  \\
  SLAC National Accelerator Laboratory\\
  E-mail: \email{wulsin@slac.stanford.edu}
}

\abstract{
  We report a measurement of the branching fractions for
  \Bzpilnu and \Bzrholnu decays, using charged and neutral $B$ decays with
  isospin constraints.  We find 
  $\BRBzpilnu  = (1.41 \pm 0.05 \pm 0.07) \times 10^{-4}$, and  
  $\BRBzrholnu = (1.75 \pm 0.15 \pm 0.27) \times 10^{-4}$, 
  where the first error is statistical and the second is systematic.  
  We measure \dBdq2, with 6 \q2
  bins for \Bzpilnu and 3 \q2 bins for \Bzrholnu, and compare the
  distributions 
  in data with theoretical predictions for the form factors.  We 
  use these branching fractions and form-factor calculations to
  determine \Vub.  Based on a combined fit to the FNAL/MILC
  lattice QCD calculation and 
  data over the full \q2 range, we find
  \Vub =  $(2.95 \pm 0.31) \times 10^{-3}$.  
}

\FullConference{35th International Conference of High Energy Physics - ICHEP2010,\\
		July 22-28, 2010\\
		Paris France}

\begin{document}

\section{Motivation}
The CKM matrix element \Vub is best determined by measuring the decay rate
for \BXulnu, which is proportional to $\Vub^2$.  The advantage
of charmless semileptonic decays over charmless hadronic $B$ decays is
that the leptonic and hadronic currents of the amplitude factorize.
Calculations of the hadronic current are difficult, since they must take
into account physical mesons rather than free quarks.  So the hadronic
current is typically parameterized by form factors which can be
calculated in the framework of QCD.  

In this analysis \cite{BAD2034} \Vub is determined with 
exclusive decays which, compared with inclusive decays, have the
advantage of reduced backgrounds but suffer from lower signal yields.    
We measure \dBdq2, the partial branching
fraction with respect to \q2, the momentum-transfer squared, for four
decay modes: 
\Bzpilnu,
\Bppizlnu,
\Bzrholnu, and
\Bprhozlnu.

\section{Data set and candidate selection}
This analysis is based on a data set of 377 million \BBbar pairs
recorded with the \babar~detector  \cite{NIM} at the PEP-II
energy-asymmetric 
\epem collider operating at the \FourS resonance.  An additional
sample of 35.1 \invfb of data was
collected at 40 \mev below the \FourS resonance, which is used to
study non-\BBbar backgrounds.  Monte Carlo (MC) techniques
\cite{evtgen} are used to simulate \BBbar physics production and
decay as well as detector efficiencies and resolutions.  

Signal $B$ candidates are selected based on their three decay products:
a high-momentum lepton ($\ell = e,
\mu$), a hadron ($\pi^\pm, \pi^0, \rho^\pm, \rho^0$), and a neutrino.  
The neutrino is reconstructed from the missing energy and momentum in the
event, and several criteria require that it be consistent with a
physical neutrino.  
The second $B$ in the event is not explicitly reconstructed; this
untagged approach increases the signal yield but results in larger
backgrounds.   

The major challenge of this analysis is to separate the relatively
small signal from the much larger backgrounds.   The largest
background comes from other \BBbar decays, in 
particular charmed semileptonic \BXclnu decays, which have a rate 
about 50 times that of charmless semileptonic \BXulnu decays.
Backgrounds also originate from \qq~($q = u,d,s,c$) events,  
which differ from \BBbar events in that they have a more jet-like
topology. 
\BXulnu decays are
very similar to the signal, and are especially prevalent at high 
\q2. 
The background composition changes as a function of \q2, with
\qq~contributing mostly at low and high \q2, \BXclnu the dominant 
background at medium \q2, and \BXulnu the largest background at
high \q2.  

To reduce background contamination, neural nets are used with inputs
based on event shape and neutrino kinematics.  Nets are trained 
separately in different ranges of \q2 against each of the three
dominant backgrounds:  \qq, \BXclnu, and \BXulnu.  

Detailed control sample studies are used to check the agreement
between simulation and data. 
The data and MC agree well in \BXclnu-enhanced samples
and exclusively reconstructed \Bzdslnu decays, suggesting that the neutrino 
reconstruction is well-modeled in the simulation.

\section{Signal yield fit and \dBdq2}
The signal yield is extracted with an extended binned maximum likelihood
fit \cite{Barlow} that takes into account the statistical
uncertainties of both the 
data and MC samples.  The fit is performed to the three-dimensional
distribution of \DeltaE, \mES,
and \q2.  The variables \DeltaE and \mES test the consistency of the
candidate with a $B$ decay, and are defined as:  
$\Delta E = E^\ast_B - \sqrt{s}/2$
and 
$\mES = \sqrt{ (\sqrt{s}/2)^2 - p^{\ast2}_B }$, 
where $\sqrt{s}$ is the center-of-mass energy of the colliding beams,
and $E^\ast_B$ and $p^\ast_B$ are the center-of-mass energy and
momentum of the recontructed $B$.
\commenthww{
A two-dimensional distribution in \DeltaE and \mES is used because the
two variables are correlated in the signal and some of the
backgrounds, and it would be difficult to determine reliable analytic
expressions for these distributions.  The shapes of the
two-dimensional \DeltaE-\mES distributions are taken from simulation
of signal and background sources in each \q2 bin.  
}
The momentum-transfer squared, \q2, is calculated as the
square of the sum of the lepton and neutrino 4-momenta.  To improve
the \q2 resolution, the neutrino momentum
is scaled to set \DeltaE to zero in the calculation of \q2.   

The four modes are fit simultaneously with isospin
constraints for the charged and neutral modes.  The fit yields $10604
\pm 376$ signal \Bpilnu decays and $3332 \pm 286$ signal \Brholnu
decays, from which the branching fractions are calculated:  
\begin{eqnarray}
  \BRBzpilnu  &=& (1.41 \pm 0.05 \pm 0.07) \times 10^{-4} \ , \nonumber \\
  \BRBzrholnu &=& (1.75 \pm 0.15 \pm 0.27) \times 10^{-4} \ , \nonumber
\end{eqnarray}
where the first error is statistical and the second error is
systematic.  Separate measurements of the branching fractions from
single-mode fits 
to charged or neutral $B$ samples are found to be consistent within
statistical uncertainties with the result from the combined four-mode
fit.  

The error on each of these branching fractions is dominated by
systematic uncertainties.  For \Bpilnu, the largest systematic
uncertainties arise from the spectrum of $K_L$, which carry away
missing energy and momentum and thus impact the neutrino resolution;
and from the shape of the fit distributions for the \qq\ background.  For
\Brholnu, the largest uncertainties arise from the shape function
parameters and branching ratio of non-resonant \BXulnu
decays, which cannot be easily distinguished from the signal.

\begin{figure}[hbt]
  \centering
  \begin{minipage}{0.45\linewidth}
    \epsfig{file=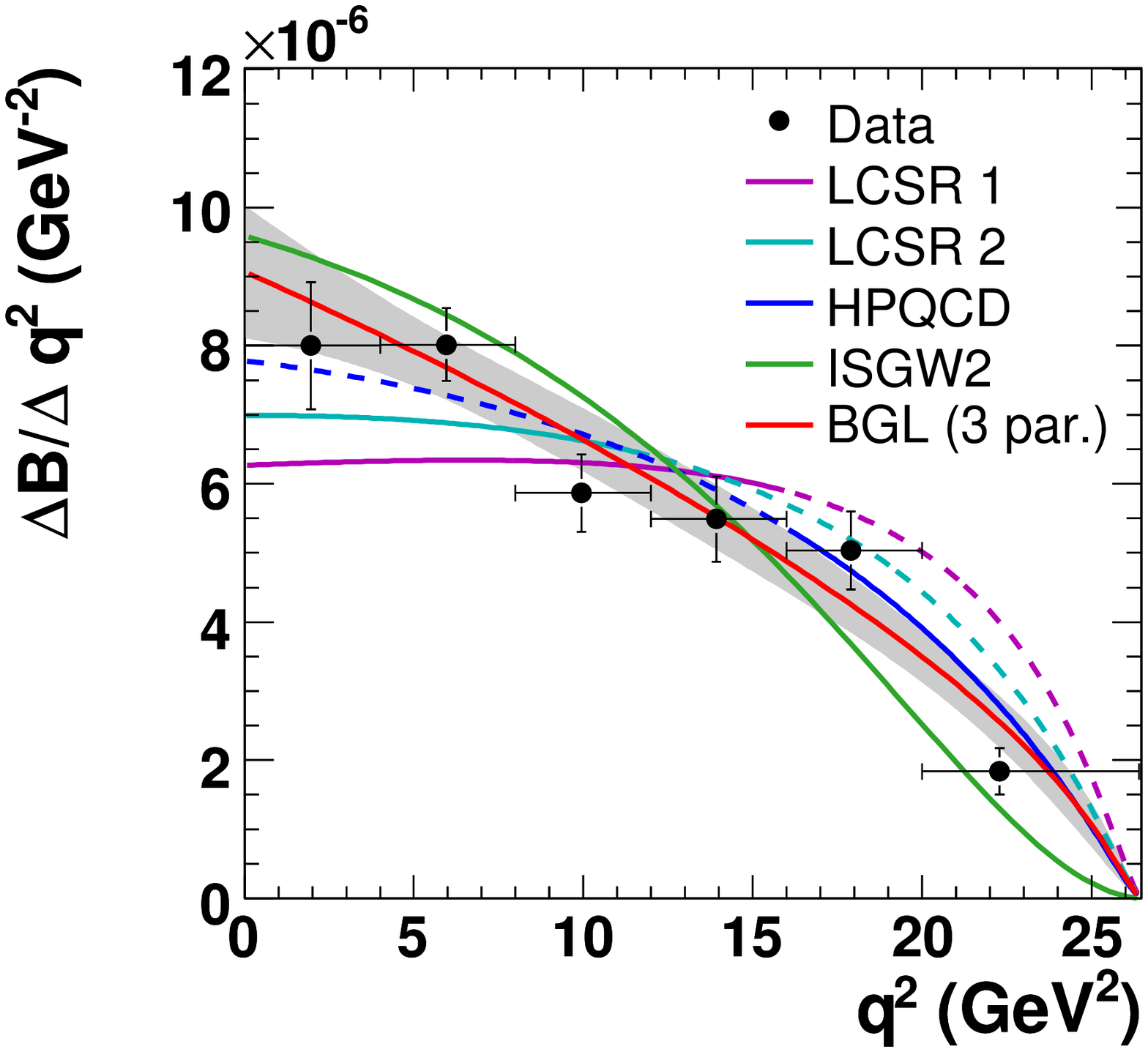, width = 7cm}\\
  \end{minipage}
  \begin{minipage}{0.45\linewidth}
    \epsfig{file=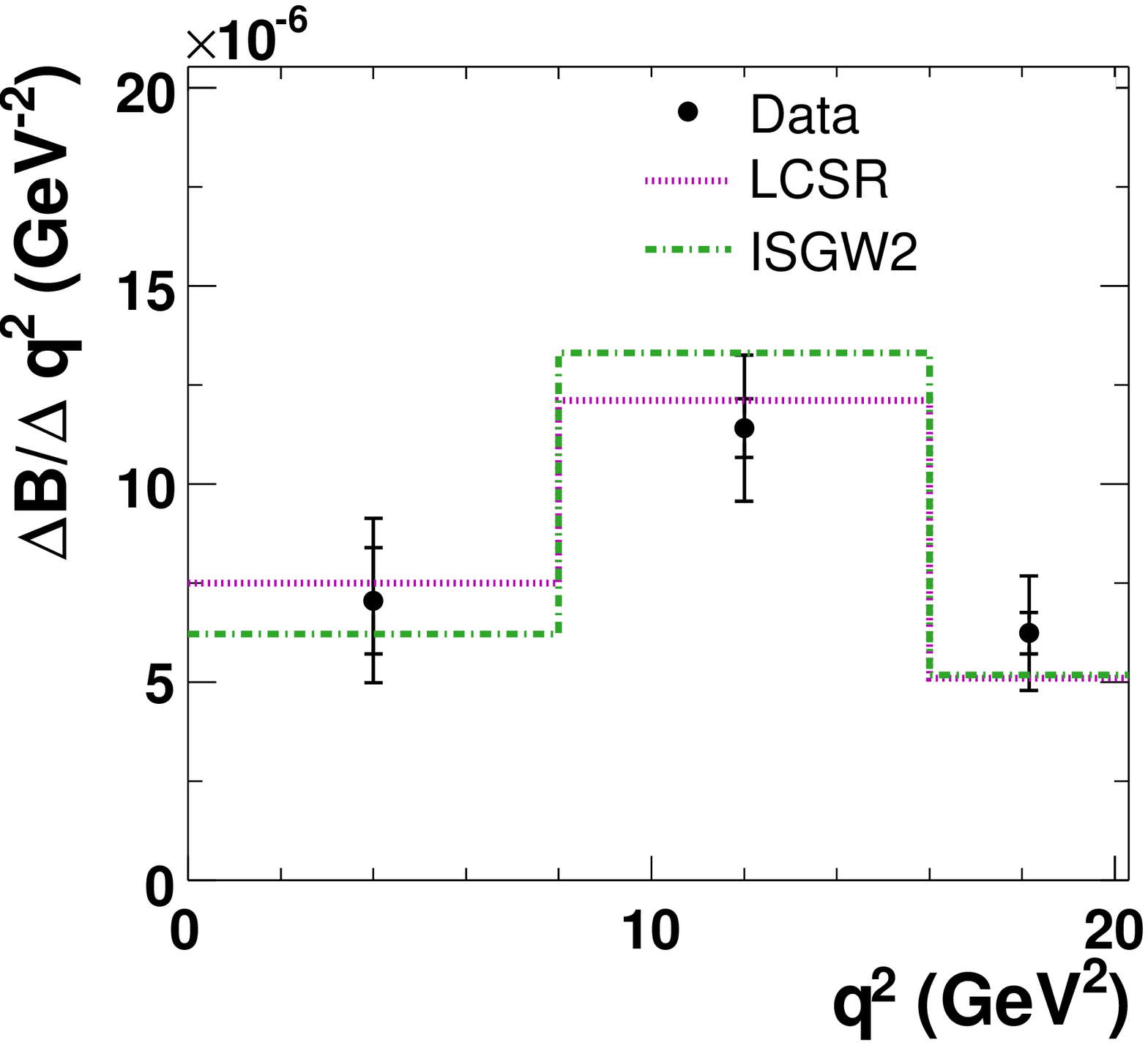, width = 7cm}\\
  \end{minipage}
  \caption{Partial branching fraction of 
    \Bzpilnu (left), 
    \Bzrholnu (right), 
    in bins of \q2.  
    Values measured from data (points with errors) are compared with
    predictions from light cone sum rules (LCSR1 \cite{LCSR-pi} and LCSR2
    \cite{LCSR2} for \Bzpilnu, and LCSR \cite{LCSR-rho} for \Bzrholnu),
    unquenched lattice QCD (HPQCD \cite{HPQCD}), and quark-model (ISGW2
    \cite{ISGW2}) theory calculations, normalized to equal area.  The BGL
    parameterization \cite{BGL} is shown with the 
    shaded region indicating the uncertainty on the fitted parameters.  
    For \Bzpilnu, the dashed lines indicate an extrapolation of the 
    prediction to the full \q2 range.
  }
  \label{fig:BF}
\end{figure}

The \Bpilnu signal
yield is fit in 6 bins of \q2, and the \Brholnu yield is measured in 3
bins of \q2, as shown in Figure \ref{fig:BF}, after correction for
efficiency, detector resolution, bremsstrahlung, and radiative
effects.  
The measured \dBdq2 distribution is compared with several theoretical
form-factor calculations.  For \Bzpilnu the agreement
with the data is best for the HPQCD 
lattice calculation \cite{HPQCD}, but the lattice QCD predictions are
only valid for $q^2 > 16 \gev^2$, the earlier light cone sum rules
(LCSR) calculation  
(LCSR~1 \cite{LCSR-pi}) for $q^2 < 16 \gev^2$, and the more recent LCSR calculation
(LCSR~2 \cite{LCSR2})   for $q^2 < 12 \gev^2$.  For
\Bzrholnu, the branching fraction measurements are not precise enough 
to discriminate between the predictions from LCSR
\cite{LCSR-rho} and the ISGW2 quark-model calculation \cite{ISGW2}. 

\section{Determination of  \Vub}
The value of \Vub can be determined from the measured partial branching
fraction over a limited range of \q2 and an integral of the form
factor over  the same range of \q2: 

\begin{equation}
  \Vub = \sqrt{\frac{\Delta{\cal B}(q^2_{min},q^2_{max})}
    {\tau_0 \ \Delta\zeta(q^2_{min},q^2_{max})}} \ , 
  \hspace{1.5cm}
  \Delta \zeta(q^2_{min},q^2_{max})  = 
  \frac{G^2_F} {24 \pi^3} 
  \int\limits_{q^2_{min}}^{q^2_{max}} 
  p_{\pi}^3 |f_+(q^2)|^2 dq^2 \ ,  \nonumber 
  \label{eq:Vub}
\end{equation}

where $\tau_0 = (1.530 \pm 0.009)$~ps is the $B^0$~lifetime.

\Vub is determined from form-factor theoretical predictions 
from light cone sum rules at low \q2 and lattice QCD at
high \q2, and in each case the
theoretical uncertainty of about 15\% dominates.  
These measurements of \Vub are shown in the first three lines of Table
\ref{tab:vub}.

\begin{table}[hbt] 
  \centering
  \begin{tabular}{llccl} \hline\hline
    & $q^2$ Range & $\Delta {\cal B}$ & $\Delta\zeta$ &
    \multicolumn{1}{c}{$|V_{ub}|$} \\
    & ($\gev^2$) & (10$^{-4}$) & (ps$^{-1}$) &
    \multicolumn{1}{c}{(10$^{-3}$)} \\
    \hline
    LCSR~1~\cite{LCSR-pi}  & $0-16$    & $1.10 \pm 0.07$ & $5.44{\pm}1.43$        & $3.63 \pm 0.12^{+0.59}_{-0.40}$ \\
    LCSR~2~\cite{LCSR2}    & $0-12$    & $0.88 \pm 0.06$ & $4.00^{+1.01}_{-0.95}$ & $3.78 \pm 0.13^{+0.55}_{-0.40}$ \\
    HPQCD~\cite{HPQCD}     & $16-26.4$ & $0.32 \pm 0.03$ & $2.02{\pm}0.55$        & $3.21 \pm 0.17^{+0.55}_{-0.36}$ \\ 
    \hline
    FNAL/MILC~\cite{FNAL}  & $0-26.4$  & $1.41 \pm 0.09$ & ---                    & $2.95 \pm 0.31$                 \\ 
    HPQCD~\cite{HPQCD}     & $0-26.4$  & $1.41 \pm 0.09$ & ---                    & $2.99 \pm 0.35$                 \\ 
    \hline\hline
  \end{tabular}
  \caption{$|V_{ub}|$ derived from \Bpilnu  decays for 
    various $q^2$ regions and form-factor calculations.
    The first three values use data points from only a limited \q2
    range, while the final two values use data from the
    full \q2 range.  
    Experimental and theoretical errors on \Vub are listed separately 
    for partial-\q2 determinations and combined into 
    a single error for the full-\q2 determinations.  
  }
  \label{tab:vub}
\end{table}

\begin{figure}[hbt]
  \centering
  \begin{minipage}{0.45\linewidth}
    \epsfig{file=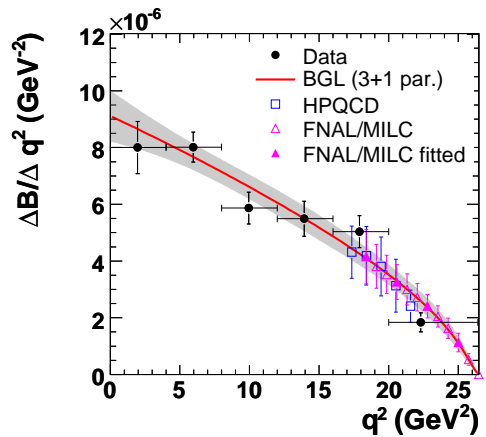, width = 7cm}\\
  \end{minipage}
  \caption{Partial branching fraction of \Bzpilnu in bins of \q2, 
    with the combined fit to data (black points) and lattice theory
    (squares or triangles) represented by the solid red line.  
    Not all theory points are
    used because of large correlations. 
    The shaded region indicates the uncertainty on the fitted parameters.  
  }  
  \label{fig:combinedFit}
\end{figure}

\Vub can also be determined using a combined fit to data and
theoretical predictions over the full range of \q2.  
Three fit parameters describe a quadratic
polynomial in the BGL form-factor ansatz \cite{BGL}, and a
fourth parameter defines 
the relative normalization of theory and data, which is
proportional to \Vub. 
The advantage of this method is that it reduces the
theoretical uncertainty.  
The relative error contributions are $3\%$ from the
branching-fraction measurement, 
$5\%$ from the shape of the $q^2$ spectrum determined from data, and
$8.5\%$  
from the form-factor normalization obtained from theory.
The result of this fit is shown in Figure
\ref{fig:combinedFit}, and the values of \Vub for two lattice
form-factor calculations are 
shown in the last two lines of Table \ref{tab:vub}.  
Unfortunately, the lattice calculations are only valid at high \q2,
where the decay rate is lowest and the experimental uncertainties are
largest.

In summary, \Vub has been determined with two methods.  
Using data from a limited \q2 range, larger values of \Vub
are found with theory predictions for low \q2 than with predictions for
high \q2. 
A combined fit to theory and data from the full \q2 range
reduces the overall error by a factor of two.  
The value of \Vub from both methods is 
smaller than most determinations of \Vub based on inclusive \BXulnu
decays, which are in the range $(4.0 - 4.5) \times 10^{-3}$.  

\clearpage


\begin{thebibliography}{99}
  
\bibitem{BAD2034}
  P.~del Amo Sanchez {\it et al.}  [\babar Collaboration],
  arXiv:1005.3288 [hep-ex], 
  accepted for publication by Phys. Rev. D.  
  
\bibitem{NIM}
  \babar\ Collaboration, B.~Aubert {\it et al.},
  \newblock Nucl. Instr. and Methods {\bf A479}, 1 (2002).
  
\bibitem{evtgen}
D. J. Lange, 
\newblock Nucl. Instr. and Methods {\bf A462}, 152 (2001).

\bibitem{Barlow}
  R.~J. Barlow and C.~Beeston,
\newblock Comput. Phys. Commun. {\bf 77}, 219--228 (1993).

\bibitem{LCSR-pi} 
  P.~Ball and R.~Zwicky,
  \newblock JHEP {\bf 0110}, 019 (2001);
  \newblock Phys. Rev. {\bf D71}, 014015 (2005).
  
\bibitem{LCSR2}
   G.~Duplancic, A.~Khodjamirian, T.~Mannel, B.~Melic and N.~Offen,
   \newblock JHEP {\bf 804}, 14 (2008).

\bibitem{LCSR-rho} 
  P.~Ball and V.~M. Braun,
  \newblock Phys. Rev. {\bf D58}, 094016 (1998);
  P.~Ball and R.~Zwicky,
  \newblock Phys. Rev. {\bf D71}, 014029 (2005).  
  
\bibitem{HPQCD} 
  HPQCD Collaboration, 
  E. Gulez, \etal, Phys. Rev. {\bf D73}, 074502 (2006) and
  Erratum ibid. {\bf D75}, 119906 (2007).
  
\bibitem{ISGW2} 
  N.\ Isgur, D.\ Scora, B.\ Grinstein, and M. B.\ Wise, 
  Phys.\ Rev.\ {\bf D39}, 799 (1989); 
  D.\ Scora, N.~Isgur, Phys.\ Rev.\ {\bf D52}, 2783 (1995).
  
\bibitem{BGL}
  C. G. Boyd, B. Grinstein, and R.F. Lebed, 
  Phys.\ Rev.\ Lett.\ {\bf 74}, 4603 (1995);
  C.G. Boyd and M.J. Savage, 
  Phys.\ Rev. {\bf D56}, 303 (1997).
  
\bibitem{FNAL}
  Fermilab Lattice and MILC Collaboration, 
  J. Bailey \etal, 
  Phys.\ Rev.\ {\bf D79}, 054507 (2009).

\end{thebibliography}
\end{document}